\documentclass{article}

\usepackage{PRIMEarxiv}

\usepackage[utf8]{inputenc} 
\usepackage[T1]{fontenc}    
\usepackage{hyperref}       
\usepackage{url}            
\usepackage{booktabs}       
\usepackage{amsfonts}       
\usepackage{nicefrac}       
\usepackage{microtype}      
\usepackage{lipsum}
\usepackage{fancyhdr} 
\usepackage{amsmath} 
\usepackage{graphicx}       
\graphicspath{{media/}}     

\title{Evaluating Imputation Techniques for Short-Term Gaps in Heart Rate Data
}

\author{
  Vaibhav Gupta \\
 Data Engineering\\
  Helmut-Schmidt-University \\
  Hamburg\\
  \texttt{guptav@hsu-hh.de} \\
   \And
  Maria Maleshkova \\
  Data Engineering\\
  Helmut-Schmidt-University \\
  Hamburg\\
  \texttt{maleshkm@hsu-hh.de} \\
}

\begin{document}
\maketitle

\begin{abstract}
Recent advances in wearable technology have enabled the continuous monitoring of vital physiological signals, essential for predictive modeling and early detection of extreme physiological events. Among these physiological signals, heart rate (HR) plays a central role, as it is widely used in monitoring and managing cardiovascular conditions and detecting extreme physiological events such as hypoglycemia. However, data from wearable devices often suffer from missing values. To address this issue, recent studies have employed various imputation techniques. Traditionally, the effectiveness of these methods has been evaluated using predictive accuracy metrics such as RMSE, MAPE, and MAE, which assess numerical proximity to the original data. While informative, these metrics fail to capture the complex statistical structure inherent in physiological signals. This study bridges this gap by presenting a comprehensive evaluation of four statistical imputation methods, Linear Interpolation, K-Nearest Neighbors (KNN), Piecewise Cubic Hermite Interpolating Polynomial (PCHIP), and B-splines, for short-term HR data gaps. We assess their performance using both predictive accuracy metrics and statistical distance measures, including the Cohen’s Distance Test (CDT) and Jensen-Shannon Distance (JS Distance), applied to HR data from the D1NAMO dataset and the BIG IDEAs Lab Glycemic Variability and Wearable Device dataset. The analysis reveals limitations in existing imputation approaches and the absence of a robust framework for evaluating imputation quality in physiological signals. Finally, this study proposes a foundational framework to develop a composite evaluation metric to assess imputation performance.
\end{abstract}


\section{Introduction}
\label{intro}

Clinical datasets used for training predictive models often suffer from missing data, a challenge stemming from imperfect monitoring of patient vital signs, sensor limitations, or intermittent physiological signal sampling \cite{Ref3}. These missing values can appear in various gap sizes, from short intervals like 5 to 30 minutes to long-term gaps exceeding 2 hours. Notably, short-term gaps are more frequent, often arising from sensor malfunctions or data loss during preprocessing \cite{Ref14}. One prominent approach to address this issue involves using appropriate imputation techniques. A review of existing literature reveals that statistical imputation methods are most frequently applied to short-term missing intervals, primarily due to their simplicity and low computational overhead \cite{Ref1}, \cite{Ref9}.

An important task in imputing missing data is evaluating the performance of imputation models \cite{Ref3}. Imprecise imputation techniques can produce misleading instances that can impact the performance of predictive models for clinical datasets \cite{Ref11}. Following is one of the motivating scenarios for our study that demonstrates the importance of evaluating imputation techniques.

Developing accurate models to predict hypoglycemia is an important focus in managing Type 1 Diabetes (T1D), a chronic autoimmune disease caused by the destruction of pancreatic beta-cells, leading to insufficient insulin production. While insulin therapy is necessary to control blood glucose levels, it also increases the risk of hypoglycemia (blood glucose below 70 mg/dL) \cite{Ref12}. Researchers use datasets containing signals from wearable devices, such as electrocardiogram (ECG), photoplethysmogram (PPG), and glucose levels, to train machine learning models for hypoglycemia prediction. Features like heart rate (HR) derived from ECG and PPG are particularly valuable for this prediction task \cite{Ref13}. However, these datasets often contain missing values, negatively affecting model predictive performance. Current studies frequently employ linear imputation to fill missing data gaps, primarily due to its simplicity and computational efficiency. While this method often yields favorable performance in terms of RMSE, it tends to oversimplify the imputation process by replacing missing segments with straight-line estimates. As a result, it fails to accurately reconstruct the complex, nonlinear dynamics characteristic of physiological signals such as HR data. This can be particularly problematic in segments where individuals experience sudden or extreme physiological events. Also, this approach fails to consider the subtle influence that various missing data scenarios can have on the performance of ML models \cite{Ref11}.

Current state-of-the-art studies predominantly use evaluation metrics such as RMSE, Mean Absolute Percentage Error (MAPE), and Mean Absolute Error (MAE) to assess the performance of imputation techniques for physiological signals. While these metrics help quantify the point-to-point proximity between original and imputed values, they primarily capture numerical accuracy and overlook the preservation of underlying data distributions.

To address the limitations of existing evaluation criteria for assessing the performance of imputation techniques, we make the following key contributions to advance a more comprehensive analysis framework. First, the paper presents a survey and analysis of the available statistical imputation techniques for imputing HR values for short-term gaps. Second, we comprehensively evaluate HR imputation techniques using two datasets, assessing their performance across two categories of state-of-the-art evaluation metrics. Third, we lay out a foundation for integrating multiple metrics to enhance the robustness of imputation performance.

The remainder of this paper is structured as follows.
Section \ref{Related work} reports on the state of the art of imputation methods and evaluation metrics on HR signals. Section \ref{methods} introduces the methodology and discusses our study's imputation and evaluation metrics. The comparative analysis of the evaluation metrics is reported in section \ref{analysis} and discussed in section \ref{discussion}. Finally, key findings are summarized in section \ref{conclusion}.

\section{Literature Review}
\label{Related work}

The application of imputation techniques for handling missing data has seen significant growth in recent years. However, the imputation techniques used to impute HR data in datasets used for hypoglycemia prediction have been limited. For example, Mantena et al. \cite{Ref8} employed K-Nearest Neighbors (KNN)-based imputation. In contrast, Vahedi et al. \cite{Ref9} and Leutheuser et al. \cite{Ref6} adopted the approach of mean substitution to fill missing HR values. In another notable instance, Bertachi et al. \cite{Ref10} used linear interpolation for gaps of up to two hours, choosing to discard data with longer missing durations.

As HR imputation research has evolved, a more diverse variety of imputation techniques has emerged, enabled by access to more comprehensive and diverse datasets. This progression is evident in the work of Mochurad et al. \cite{Ref1} and Lin et al. \cite{Ref2}, who employed both statistical and deep learning-based imputation strategies. Mochurad et al. \cite{Ref1} implemented Bezier and B-spline interpolation techniques. In contrast, Lin et al. \cite{Ref2} explored a variety of gap durations (2, 4, 6, and 8 hours within a 24-hour window), applying traditional methods such as linear interpolation, exponentially weighted moving average, KNN, Kalman smoothing, and Last Observation Carried Forward (LOCF), alongside advanced deep learning approaches including Denoising Autoencoders, bi-directional RNNs, Context Encoders, the Spatial-Temporal Completion Network for Video Inpainting, and HeartImp. Chakrabarti et al. \cite{Ref17} used binned data for imputation and used expectation maximization (EM) algorithm, k-nearest neighbor (kNN), iterative imputer (II), random forest (RF), and the simple imputer (SI) method to impute missing heart rate data for 15 min and 1 hr of missing data.

Despite the increasing use of these imputation methods, a notable gap exists in the systematic evaluation of their effectiveness. Most studies rely on standard predictive accuracy metrics such as RMSE, MAPE, and MAE. For instance, Mochurad et al. \cite{Ref1} utilized MAPE and the coefficient of determination ($R^2$), while Lin et al. \cite{Ref2} incorporated RMSE, MAPE, and MAE in their evaluation framework. While helpful in evaluating numerical accuracy, these metrics are often insufficient to fully capture the complex temporal structures and physiological patterns inherent in HR data.

Recognizing these limitations, recent studies have begun to propose complementary evaluation criteria, but for tabular data, incorporating Histograms and statistical distance metrics like Jensen-Shannon (JS) distance and the Cohen’s Distance Test (CDT) \cite{Ref3} with RMSE scores. However, to the best of the authors' knowledge, no existing study has systematically compared the performance of statistical imputation techniques for short-term HR data gaps using both predictive accuracy and statistical distance metrics. This gap underscores the need for a more comprehensive evaluation framework considering signal fidelity and clinical relevance, especially in critical applications such as hypoglycemia prediction and personalized healthcare.

\section{Methods and Research Design}
\label{methods}
To address the key research gaps in the current state of the art, this study evaluates the performance of state-of-the-art imputation techniques based on predictive and statistical evaluation criteria. To evaluate the methodology, we use HR values of two different datasets: diabetic patients of D1NAMO \cite{Ref4} and pre-diabetic patients of BIG IDEAs Lab Glycemic Variability and Wearable Device Data \cite{Ref15} (refer to table \ref{tab-bigideas} and table \ref{tab-d1namo} for the detailed summary). As a first step, we standardize the timestamps in both datasets by converting them to Unix time, ensuring consistency for downstream processing. Then, we clean the dataset by removing missing and duplicate values from the HR data. To evaluate the imputation technique on the original HR data, we create artificial gaps of 5 and 15 minutes separately. These gaps are inserted sequentially throughout the time series with a 1-minute interval between each gap. These missing segments are subsequently imputed using statistical imputation techniques, including Linear Interpolation, Piecewise Cubic Hermite Interpolating Polynomial (PCHIP), K-Nearest Neighbors (KNN), and B-Spline Interpolation. 
The performance of each technique over the individual imputed gap is evaluated using two categories of state-of-the-art evaluation metrics: three predictive accuracy and two statistical distance metrics. The overall performance of each imputation technique is calculated by averaging the evaluation metric scores across all gaps. This is done separately for each gap size, dataset, and imputation technique.

\begin{table}[h!]
\centering
\caption{Summary of BIG IDEAs Lab Glycemic Variability and Wearable Device Data}
\label{tab-bigideas}
\begin{tabular}{ll}
\hline
\textbf{Dataset} & \textbf{BIG IDEAs} \\\hline
\textbf{Subjects} & 16 (Pre-Diabetes patients) \\
\textbf{Age Range} & 35--65 years \\
\textbf{A1C Measurement} & 5.2\% - 6.4\% \\
\textbf{CGM Device} & Dexcom G6 \\
\textbf{Sensor Device} & Empatica E4 wristband \\
\textbf{Study Duration} & 10 days \\\hline
\end{tabular}

\end{table}

\begin{table}[h!]
\centering
\caption{Summary of D1NAMO Dataset}
\label{tab-d1namo}
\begin{tabular}{ll}
\hline
\textbf{Dataset} & \textbf{D1NAMO} \\\hline
\textbf{Subjects} & 20 healthy, 9 T1D patients \\
\textbf{CGM Device} & iPro2 Professional CGM sensor \\
\textbf{Sensor Device} & Zephyr Bioharness 3 sensor \\
\textbf{Signals recording} & 450h \\
\textbf{Glucose measurements} & 8414 \\
\textbf{Food pictures} & 106 \\\hline
\end{tabular}
\end{table}

The description of the imputation techniques and evaluation metrics employed are discussed in detail in the following subsections. 

\subsection{Statistical Imputation Techniques}
Statistical imputation techniques are methods used to estimate and replace missing data values based on observed data patterns. These techniques leverage statistical models or equations to preserve the integrity and structure of the original dataset. Several imputation techniques are commonly employed to address short-term gaps in HR data. These include Linear, PCHIP, B-spline, and KNN, each described as follows.

\begin{itemize}
    \item Linear interpolation estimates missing values by connecting two known points with a straight line, assuming a constant rate of change.
    \item KNN imputes missing data by identifying the k most similar data points and computing a weighted average, where k is a tunable parameter. In our study, we have used k = 5. 
    \item The process of B-Spline interpolation involves constructing B-Spline basis functions \( B_{i,k}(x) \), where \( k \) is the degree, satisfying the conditions: 1) Each basis function \( B_{i,k}(x) \) is defined over the interval \([t_i, t_{i+k+1}]\); 2) \( B_{i,k}(x) \) is a polynomial of degree \( k \) within each interval \([t_i, t_{i+1}]\); 3) Each point \( x \) lies within at most \( k+1 \) neighboring basis functions. Here, a polynomial of degree 3 is implied. 
\item PCHIP maintains the monotonicity and trend of the data. It uses piecewise cubic polynomials between each pair of data points, defined as: 
\begin{equation} \label{eq2}
y(x)=\sum_{i=1}^{n-1}(a_i(x-x_i)^3+b_i(x-x_i)^2+c_i(x-x_i)+d_i)
\end{equation}
This interpolation technique is beneficial for preserving trends in the data and avoiding overshooting.
    
\end{itemize}
 
\subsection{Evaluation Metrics}
Evaluation metrics for imputation performance can generally be grouped into two main categories: Predictive accuracy metrics and Statistical distance metrics.

Predictive accuracy metrics assess the point-wise differences between imputed values and their corresponding ground truth values. Standard metrics in this category include Root Mean Square Error (RMSE), Mean Absolute Error (MAE), and Mean Absolute Percentage Error (MAPE). RMSE penalizes larger errors more heavily by squaring the differences before averaging, making it sensitive to outliers. MAE measures the average absolute difference between predicted and actual values, offering a straightforward interpretation of average error magnitude. MAPE expresses the prediction error as a percentage of the actual values, allowing for relative error comparison across different scales. These metrics are widely used in time series imputation tasks due to their simplicity, interpretability, and ability to provide clear quantitative insights. Lower values of these metrics indicate higher similarity between the imputed and actual data, signifying a more accurate reconstruction of the original time series.

Unlike predictive accuracy metrics, statistical distance metrics assess how closely the probability distributions of imputed data match those of the original data. In this study, one such statistical metric employed is Cohen’s Distance Test (CDT), a standardized measure of effect size that quantifies the difference between two group means in terms of their pooled standard deviation. It is beneficial for evaluating the similarity between the distributions of actual and imputed HR values. The CDT is defined in Eq. \ref{cd}, where $\tilde{x}$ and $SD$ are the mean and standard deviations of the actual and imputed distributions. Distributions with small, medium, and significant differences have a CDT $\leq 0.2$, $0.2 < \text{CDT} \leq 0.5$, and $0.5 < \text{CDT} \leq 0.8$, respectively \cite{Ref18}. 
\begin{equation}
\label{cd}
CDT = \frac{\tilde{x}_R - \tilde{x}_I}{SD_p}
\quad \text{where} \quad
SD_p = \sqrt{\frac{SD_R^2 + SD_I^2}{2}}
\end{equation}

Our study's other statistical distance metric is Jensen-Shannon Distance (JSD). It is a symmetric and smoothed version of the Kullback-Leibler (KL) divergence, used to measure the similarity between two probability distributions (refer \ref{jsd}). A JSDist $= 0$ indicates identical distributions, while JSDist $= 1$ represents maximally different distributions \cite{Ref19}.
\begin{equation}
\label{jsd}
JSDist = \sqrt{ \frac{KL(p, r)}{2} + \frac{KL(q, r)}{2} }
\quad \text{where}
\quad KL(a, b) = \sum_{i=0}^{a_{\text{bins}}} a_i \cdot \log_{2} \left( \frac{a_i}{b_i} \right),
\quad r = \frac{p + q}{2}
\end{equation}

\section{Comparative Analysis}
\label{analysis}

According to the methodology described, in this section, we compare the performance of the imputation techniques. 
Our analysis is divided into subparts- based on predictive and statistical distance metrics.

\subsection{Predictive Metric Analysis}
Based on predictive accuracy metrics, performance was evaluated for gap sizes of 5 and 15 minutes across both datasets (refer Tables \ref{tab-1}, \ref{tab-2}, \ref{tab-3} and \ref{tab-4}). In terms of RMSE, the Linear imputation method outperformed other techniques, achieving the lowest errors of 11.45 and 13.27 for 5-minute and 15-minute gaps, respectively, on the D1NAMO dataset and 10.22 and 12.25 on the BIG IDEAs Lab Glycemic Variability and Wearable Device Data.

Similarly, when evaluated using MAE, Linear again showed superior performance, with scores of 10.08 and 11.41 for 5-minute and 15-minute gaps on D1NAMO and 9.21 and 10.65 for the corresponding gaps on the BIG IDEAs Lab Glycemic Variability and Wearable Device Data. However, considering the MAPE, the results differ. On the D1NAMO dataset, PCHIP achieved the best MAPE scores of 17.05 and 19.56 for 5-minute and 15-minute gaps, respectively. In contrast, for the BIG IDEAs Lab Glycemic Variability and Wearable Device Data, Linear again led with MAPE values of 10.84 and 12.54.

\subsection{Statistical Distance Metric Analysis}

Based on Statistical distance metrics evaluation, for the D1NAMO dataset, Linear imputation consistently achieved the lowest CDT scores of 0.542 and 0.522 for 5-minute and 15-minute gaps, suggesting it best preserved the original HR distribution's mean and standard deviation structure. However, PCHIP outperformed others in JS Distance with scores of 0.817 and 0.793, indicating it more closely captured the overall distributional shape of the original data (refer Tables \ref{tab-1} and \ref{tab-2}).

In contrast, for the BIG IDEAs Lab Glycemic Variability and Wearable Device Data, PCHIP achieved the best CDT scores at 5-minute and 15-minute gaps of 0.660 and 0.611. At the same time, Linear imputation showed the lowest JS Distance with scores of 0.790 and 0.767. This reversal suggests that PCHIP better approximated distributional statistics in this dataset, whereas Linear maintained a closer probabilistic overlap with the original data (refer Tables \ref{tab-3} and \ref{tab-4}).

\begin{table}
\centering
\caption{Analysis of Imputation techniques for HR values of D1NAMO dataset for 5 min gaps}
\label{tab-1}       
\begin{tabular}{lllll}
\hline
 &Linear & PCHIP & B spline & KNN \\\hline
RMSE& \textbf{11.45} & 11.60 & 16.69 &  14.71\\
MAE& \textbf{10.08} & 10.18 & 15.29 & 13.36\\
MAPE & 17.07 & \textbf{17.05} & 24.54 & 21.57 \\
CDT & \textbf{0.542} & 0.921 & 0.921 & 0.927\\
JS distance & 0.822 & \textbf{0.817} & 0.866 & 0.891\\\hline
\end{tabular}
\end{table}  

\begin{table}
\centering
\caption{Analysis of Imputation techniques for HR values of D1NAMO dataset for 15 min gaps}
\label{tab-2}       
\begin{tabular}{lllll}
\hline
 &Linear & PCHIP & B spline & KNN \\\hline
RMSE& \textbf{13.27} & 13.50 & 18.20 & 16.38\\
MAE& \textbf{11.41 }& 11.60 & 16.35 &14.55\\
MAPE & 19.60 & \textbf{19.56} & 26.45 &23.71\\
 CDT &\textbf{0.522}  & 0.534 & 0.846 &0.855\\
 JS distance& 0.795 & \textbf{0.793} & 0.861&0.889\\\hline
\end{tabular}
\end{table}

\begin{table}
\centering
\caption{Analysis of Imputation techniques for HR values of BIG IDEAs Lab Glycemic Variability and Wearable Device Data for 5 min gaps}
\label{tab-3}       
\begin{tabular}{lllll}
\hline
 &Linear & PCHIP & B spline & KNN \\\hline
RMSE& \textbf{10.22} & 10.35 & 13.86 & 12.78\\
MAE&\textbf{ 9.21} & 9.31 & 12.77 & 11.70\\
MAPE & \textbf{10.84} & 10.89 & 15.56 &14.04 \\
 CDT & 0.686 &\textbf{0.660}  & 1.166 &1.173\\
JS distance & \textbf{0.790} & 0.80 & 0.961 &0.982\\\hline
\end{tabular}
\end{table} 

\begin{table}
\centering
\caption{Analysis of Imputation techniques for HR values of BIG IDEAs Lab Glycemic Variability and Wearable Device Data for 15 min gaps}
\label{tab-4}       
\begin{tabular}{lllll}
\hline
 &Linear & PCHIP & B spline & KNN \\\hline
RMSE& \textbf{12.25} & 12.53 & 15.66 & 14.58\\
MAE& \textbf{10.65} & 10.90 & 14.08 &13.01 \\
MAPE & \textbf{12.54} & 12.73 &17.12 &15.63 \\
 CDT &0.612 &\textbf{0.611}  & 1.048 &1.047\\
JS distance & \textbf{0.769} & 0.782 & 0.960 &0.977\\\hline
\end{tabular}
\end{table}

\section{Discussion}
\label{discussion}

Datasets used for hypoglycemia prediction often suffer from missing data, which undermines statistical validity, contributes to class imbalance, and degrades model performance, particularly when gaps occur during critical windows, such as the 30 minutes preceding an event. While simple statistical imputation methods like linear interpolation are commonly applied in diabetes management due to their efficiency and ease of use for short gaps, they often fail to capture the complex, non-linear temporal dynamics of physiological signals. Imputing missing data has been shown to improve predictive performance in clinical applications, including glycemic control and hypoglycemia detection~\cite{Ref11}.

HR, a key feature for predicting hypoglycemia~\cite{Ref13}, exemplifies this challenge due to its inherently irregular and non-linear behavior. HR signals can exhibit rapid fluctuations, sharp peaks and troughs, driven by external and internal factors such as physical activity, emotional stress, and metabolic changes. For instance, HR can spike from 80 bpm to 120 bpm within a mere five minutes of exertion. Traditional imputation methods, which tend to rely on smoothed trends, often fail to replicate such abrupt variations, leading to flattened and biologically implausible imputations.

In the context of HR data imputation, predictive accuracy metrics such as RMSE, MAE, and MAPE offer the advantage of directly quantifying the numerical differences between imputed and actual values. However, a key limitation of these metrics is their inability to assess the structural or temporal dynamics of HR signals, an important consideration in clinical applications, where patterns such as sharp peaks or sudden drops (e.g., during physical activity or hypoglycemic episodes) convey critical physiological information. Based on the results presented in our study and evaluated through predictive metrics, linear imputation generally outperforms other methods, achieving the lowest MAE and RMSE scores in most cases, with PCHIP performing closely behind. However, in some instances of MAPE, PCHIP outperforms Linear imputation, indicating that no single method consistently dominates across all metrics. This underscores the limitation of relying solely on one predictive accuracy criteria for selecting the most effective imputation technique, while Linear may lead overall, metric-specific performance variations highlight the need for a more nuanced, context-driven evaluation.

On the other hand, statistical distance metrics like CDT and JS Distance evaluate the distributional similarity between the imputed and original data, offering a comprehensive view of how well the imputation preserves physiological patterns. This is particularly valuable for HR data, where trends and density shifts often signal underlying health events. However, these metrics can mask local inaccuracies, as they may suggest strong distributional alignment even when specific time points diverge significantly from ground truth. 
Based on the results, for the D1NAMO dataset, Linear imputation achieved the lowest CDT values, indicating strong preservation of distributional mean characteristics. At the same time, PCHIP performed best in JS Distance, capturing the overall shape of the distribution more accurately. Conversely, for the BIG IDEAs Lab Glycemic Variability and Wearable Device Data, PCHIP yielded the best CDT scores, while Linear attained the lowest JS Distance. These results highlight that no single method consistently dominates across statistical metrics or datasets. This underscores the importance of evaluating imputation performance through multiple criterions to ensure quantitative accuracy and physiological plausibility.

Through the analysis of both categories of metrics, the distinct strengths and limitations of predictive accuracy metrics (such as RMSE, MAE, and MAPE) and statistical distance metrics (such as CDT Distance and JS Distance), it becomes evident that neither category alone offers a complete picture of imputation quality, especially for complex physiological signals like HR. Predictive metrics effectively quantify point-wise errors but often overlook temporal trends and distributional fidelity, while statistical metrics capture overall shape and structure but may mask local inaccuracies. The analysis also reveals that the performance of imputation models and evaluation metrics is influenced by dataset characteristics, such as data size. Variations in model performance across different metrics highlight the dependence of imputation outcomes on these underlying dataset properties.

Therefore, to ensure that imputation methods are numerically accurate and physiologically meaningful, adopting a combined evaluation approach is essential. Such a composite metric would integrate value- and distribution-level perspectives, enabling a more balanced and reliable assessment of how well the imputed data reflects the underlying signal. This dual consideration is especially critical in clinical applications, where precision and pattern preservation are vital for informed decision-making.

\section{Conclusion and Future Work}
\label{conclusion}
In this study, we systematically reviewed and evaluated statistical imputation techniques for HR data using a range of predictive accuracy and statistical distance metrics. Our analysis revealed a significant gap in the ability of existing statistical imputation methods to effectively capture the intrinsic, dynamic patterns characteristic of physiological HR signals. Furthermore, we demonstrated that relying solely on a single evaluation criterion, whether predictive or statistical, can lead to incomplete or misleading interpretations of imputation performance. This underscores the necessity of adopting a composite evaluation framework that integrates multiple metric types to more holistically assess imputation quality.

As part of our future work, we aim to develop a robust statistical imputation technique specifically tailored to the complexities of physiological signals. In parallel, we will design a multidimensional evaluation framework that synthesizes the strengths of various existing metrics, enabling more comprehensive and context-sensitive assessments of imputation performance across diverse healthcare applications.


\begin{thebibliography}{}

\bibitem{Ref12}
American Diabetes Association Professional Practice Committee, 2. Diagnosis and Classification of Diabetes: Standards of Care in Diabetes—2024. \textit{Diabetes Care} \textbf{47}(Suppl 1), S20--S42 (2024). \url{https://doi.org/10.2337/dc24-S002  }

\bibitem{Ref1}
L. Mochurad and Y. Mochurad, Parallel algorithms for interpolation with bezier curves and b-splines for medical data recovery. In: \textit{International Workshop on Informatics \& Data-Driven Medicine} (2023).

\bibitem{Ref2}
S. Lin, X. Wu, G. Martinez, and N. V. Chawla, Filling Missing Values on Wearable-Sensory Time Series Data. In: \textit{Proceedings of the Conference on Wearable Technologies}, pp. 46--54.

\bibitem{Ref3}
O. Boursalie, R. Samavi, and T. E. Doyle, Evaluation Metrics for Deep Learning Imputation Models. In: \textit{Springer International Publishing, Cham} (2022), pp. 309--322.

\bibitem{Ref4}
F. Dubosson, J.-E. Ranvier, S. Bromuri, J.-P. Calbimonte, J. Ruiz, and M. Schumacher, The open d1namo dataset: A multi-modal dataset for research on non-invasive type 1 diabetes management. \textit{Informatics in Medicine Unlocked} \textbf{13}, 92--100 (2018). \url{https://doi.org/10.1016/j.imu.2018.09.003   }

\bibitem{Ref5}
J. Ahlert, T. Klein, F. Wichmann, and R. Geirhos, How aligned are different alignment metrics? (2024). \textit{Preprint or Conference Paper}.

\bibitem{Ref6}
H. Leutheuser, M. Bartholet, A. Marx, M. Pfister, M.-A. Burckhardt, S. Bachmann, and J. E. Vogt, Predicting risk for nocturnal hypoglycemia after physical activity in children with type 1 diabetes. \textit{Frontiers in Medicine} \textbf{11}, (Oct. 2024). \url{https://www.frontiersin.org/journals/medicine/articles/10.3389/fmed.2024.1439218  }

\bibitem{Ref7}
M. Benchekroun, B. Chevallier, V. Zalc, D. Istrate, D. Lenne, and N. Vera, The impact of missing data on heart rate variability features: A comparative study of interpolation methods for ambulatory health monitoring. \textit{IRBM} \textbf{44}, 100776 (Aug. 2023). \url{https://doi.org/10.1016/j.irbm.2023.100776 }


\bibitem{Ref8}
S. Mantena, A. Arévalo, J. Maley, S. da Silva Vieira, R. Mateo-Collado, J. da Costa Sousa, and L. Celi, Predicting hypoglycemia in critically ill patients using machine learning and electronic health records. \textit{Journal of Clinical Monitoring and Computing} \textbf{36}, 1297--1303 (2022). \url{https://doi.org/10.1007/s10877-021-00760-7 }

\bibitem{Ref9}
M. R. Vahedi, K. B. MacBride, W. Wunsik, Y. Kim, C. Fong, A. J. Padilla, M. Pourhomayoun, A. Zhong, S. Kulkarni, S. Arunachalam, and B. Jiang, Predicting glucose levels in patients with type 1 diabetes based on physiological and activity data. In: \textit{Proceedings of the 8th ACM Workshop on Mobile Health (MobileHealth’18)} (Association for Computing Machinery, New York, NY, USA, 2018).

\bibitem{Ref10}
A. Bertachi, C. Viñals, L. Biagi, I. Contreras, J. Vehí, I. Conget, and M. Giménez, Prediction of nocturnal hypoglycemia in adults with type 1 diabetes under multiple daily injections using continuous glucose monitoring and physical activity monitor. \textit{Sensors} \textbf{20}, 1705 (2020). \url{https://doi.org/10.3390/s20061705 }

\bibitem{Ref11}
N. U. Rehman, I. Contreras, A. Beneyto, and J. Vehi, The impact of missing continuous blood glucose samples on machine learning models for predicting postprandial hypoglycemia: An experimental analysis. \textit{Journal of Clinical Medicine} \textbf{9}(11), (2020). \url{https://doi.org/10.3390/math12101567 }

\bibitem{Ref13}
D. Dave, K. Vyas, K. Branan, S. McKay, D. J. DeSalvo, R. Gutierrez-Osuna, G. L. Cote, and M. Erraguntla, Detection of Hypoglycemia and Hyperglycemia Using Noninvasive Wearable Sensors: Electrocardiograms and Accelerometry. \textit{Journal of Diabetes Science and Technology} \textbf{18}(2), 351--362 (2024). \url{https://doi.org/10.1177/19322968221116393}


\bibitem{Ref14}
B. Cinar, Diadata: Daily life data of type-1 diabetes patients. \textit{Kaggle Dataset} (2022). Accessed: 2025-05-09. \url{https://www.kaggle.com/datasets/beyzacinar22/diadata}

\bibitem{Ref15}
P. Cho, J. Kim, B. Bent, and J. Dunn, BIG IDEAs Lab Glycemic Variability and Wearable Device Data (version 1.1.1). \textit{PhysioNet} (2023). \url{https://doi.org/10.13026/73s9-cw03}

\bibitem{Ref16}
P. Schreiber, S. Burbach, B. Cinar, L. Mackert, and M. Maleshkova,
”VitaStress: A multimodal dataset for stress detection,” unpublished.

\bibitem{Ref17}
S. Chakrabarti, N. Biswas, K. Karnani, V. Padul, L. D. Jones, S. Kesari, and S. Ashili, “Binned Data Provide Better Imputation of Missing Time Series Data from Wearables,” *Sensors*, vol. 23, no. 3, p. 1454, 2023. \url{https://doi.org/10.3390/s23031454}

\bibitem{Ref18}
Cohen, J. (1988). Statistical Power Analysis for the Behavioral Sciences (2nd ed.). Routledge. \url {https://doi.org/10.4324/9780203771587}

\bibitem{Ref19}
Briët, J., Harremoës, P. (2009). Properties of classical and quantum Jensen–Shannon divergence. Physical Review A, 79(5), 052311. \url{https://doi.org/10.1103/PhysRevA.79.052311}

\end{thebibliography}
\end{document}